# Inelastic electron scattering induced quantum coherence: isotope effect

**Akshay Kumar[$] and Vaibhav S. Prabhudesai[*]**

Tata Institute of Fundamental Research, Mumbai 400005 India

[$]Present Address: Space Application Centre, Indian Space Research Organisation, Ahmedabad 380015, India
*Author to whom any correspondence should be addressed.

**E-mail:** vaibhav@tifr.res.in

## Abstract

Recent discoveries of electron-induced coherence in both resonant and non-resonant interactions have introduced new perspectives in the field. In non-resonant processes, coherence has been observed in dipolar dissociation, where electron-induced excitation forms a coherent superposition of states of opposite parities, resulting in asymmetry in the angle-differential cross-section of the process relative to the incident electron beam. Notably, an isotope effect has been observed in $D_2$ at 50 eV, where heavier isotopes exhibit diminished asymmetry due to their longer dissociation times. Here, we report the isotope effect on quantum coherence in $D_2$ across different electron energies. Additionally, we investigate the role of coherence in the isotopologue HD. Our findings reveal that the asymmetric masses in HD do not influence electron-impact excitation, leading to similar asymmetry in the angular distributions of $H^-$ and $D^-$ ions. This observation is explained by the homonuclear-like behavior of HD within the Franck-Condon region.

## 1. Introduction

Electron-molecule interactions are essential processes in physics[1,2], chemistry[3,4], and materials science[5]. These interactions, which involve the scattering, absorption, or emission of electrons by molecules, underpin various natural phenomena and technological applications. Researchers investigate these interactions through a combination of experimental techniques[1,6] and theoretical models[7–10]. Recent advancements, such as momentum imaging of anions[6], have opened new frontiers in the field by enabling detailed visualization of anions formed during electron-molecule interactions. One of the striking discoveries enabled by momentum imaging is the observation of single-electron capture-induced quantum coherence in molecular dynamics[11]. The dissociative electron attachment (DEA) to homonuclear diatom, $H_2$, at around 14 eV shows forward-backward asymmetry in the angular distribution of the hydride ion formation, not expected for an inversion symmetric system. This phenomenon has been further generalized in electron scattering experiments[12], where electron-induced excitation leads to the formation of a coherent superposition of two neutral excited states. The non-resonant nature of this process makes the result more generic and points towards the possibility of coherently controlling reactions through incoherent electron collisions.

In DEA, auto-detachment of the electron from the negative ion resonance competes with dissociation. Isotopic substitution, such as replacing H with D, significantly reduces the DEA cross-section[13], as the parent anion's survival probability depends exponentially on dissociation time. This effect is evident in $H_2$, where heavier isotopologues exhibit much lower cross-sections. The isotope effect on DEA is most striking at the 14 eV peak, where the forward-backward asymmetry[11,14] in the angular distribution of $H^-$ ions contrasts the expected inversion symmetry of homonuclear diatomic molecules. This asymmetry arises from the interference of two dissociating quantum paths, initiated by electron attachment and leading to the same dissociation limit ($H(n = 2) + H^-(1S)$). The coherent superposition of two resonances with opposite parity creates the asymmetry, influenced by the phase difference and relative contributions of the paths. Isotopic substitution, such as $H_2$ to $D_2$, alters the phase difference due to longer dissociation times and shifts the amplitude of each path based on resonance lifetimes. For heavier isotopes, shorter-lived resonances contribute less, reducing the asymmetry, as observed in $D_2$ compared to $H_2$ at 14 eV. A similar effect is seen in electron scattering experiments, where isotopic substitution reduces the forward-backward asymmetry of $D^-$ ions observed at 50 eV incident electron energy.

HD, a distinct isotopologue within the $H_2$ family, differs fundamentally from its homonuclear counterparts, $H_2$ and $D_2$, due to the mass disparity between hydrogen (H) and deuterium (D). This difference in mass disrupts the inversion symmetry typically observed in homonuclear diatomic molecules like $H_2$ and $D_2$. Unlike these symmetric molecules, HD possesses a permanent dipole moment[15] that influences its physical properties. This dipole moment gives rise to a characteristic rovibrational spectrum[16], which has been extensively studied for its precision in determining fundamental physical constants. For instance, the rovibrational spectrum of HD has been used to accurately measure the electron-to-proton mass ratio[17], a critical parameter in atomic physics. Additionally, this spectrum has been employed to set stringent limits on quantum forces potentially arising from dark matter[18] interactions.

Under the Born-Oppenheimer approximation, HD behaves like a homonuclear molecule at short internuclear distances. However, at larger inter-nuclear distances, the term in the Hamiltonian that couples the electronic and nuclear degrees of freedom breaks the inversion symmetry[19] of HD. This breaking of inversion symmetry at larger distances results from the different masses of hydrogen (H) and deuterium (D), which influence the coupling between electronic and nuclear motions and lead to distinct chemical properties compared to homonuclear diatomic molecules.

In the study of two-colour photodissociation, Sheehy *et al.*[20] observed that in $HD^+$, both fragment ions ($H^+$ and $D^+$) exhibit identical inversion asymmetry about the polarization of light. This asymmetry changes with the relative phase between the two light fields but remains identical across both channels. This similarity in asymmetry is attributed to the homonuclear-like behavior of $HD^+$ in the Franck-Condon region.

In DEA experiments on HD at 10 eV, Swain *et al.*[21] observed that the behavior of negative ion resonances in HD was similar to that in $H_2$ and $D_2$ during electron attachment. Specifically, the angular distribution of $H^-$ and $D^-$ ions from HD closely resembles that of anionic fragments from $H_2$ and $D_2$. This resemblance suggests that the permanent dipole moment of HD does not significantly influence electron attachment processes under these conditions. To explore the effect of the small difference in the dissociation limits of $H^-$ and $D^-$ ions, which was observed in quantum interference at 14 eV, Swain *et al.*[22] also investigated DEA dynamics around 14 eV for HD. They found a near-identical asymmetry in the angular distributions of $H^-$ and $D^-$ ions, which was interpreted as the homonuclear-like behaviour of HD in the Franck-Condon region.

Dipolar dissociation (DD) is a charge-asymmetric dissociation of the excited molecule. It is one of the dissociation channels arising from the non-resonant inelastic scattering of particles. DD is a non-resonant process in which a molecule, after excitation to a particular state, dissociates to give an ion pair. The competing decay channel for DD is autoionization, analogous to autodetachment in DEA. On surviving the autoionization, the molecule may dissociate to give an ion pair. The rate of autoionization depends on the dissociation timescale, similar to how autodetachment depends on dissociation time in DEA. As a result, DD cross-sections also exhibit an isotope effect[13].

To investigate the influence of isotopes on quantum interference observed in DD for $H_2$, we conducted measurements for DD in $D_2$ for different electron energies. For HD, the dissociation can lead to two primary limits: $H^- + D^+$ and $H^+ + D^-$, which differ in energy by approximately 2.3 meV[23]. Understanding the impact of this energy difference on quantum interference in DD is crucial. Here, we discuss our experimental findings addressing these questions. We explore how this energy disparity manifests both generally in DD processes and specifically in the observed quantum interference patterns.

## 2. Experimental setup

The details of the experimental setup are described elsewhere[24, 25]; here, we provide a brief overview. A magnetically collimated pulsed electron beam is crossed with an effusive molecular beam generated using a capillary array. The interaction region is defined by the overlap of the electron and molecular beams, ensuring precise spatial confinement of the interaction. The interaction region of the spectrometer is positioned between two electrodes, a pusher and a puller, separated by 20 mm. Ions generated in the interaction region are extracted and accelerated toward the detector using the delayed pulsed extracting voltage on the pusher electrode. The extraction voltage pulse of 250ns duration is applied with a delay of 160 ns after the electron pulse of 100 ns duration. A four-lens assembly and a flight tube are used to velocity-focus the ions, ensuring accurate imaging at the detector. The ions are detected by a two-dimensional position-sensitive detector, which consists of a microchannel plate (MCP) arranged in a chevron configuration, coupled to a phosphor screen. The phosphor screen output is recorded by a CCD camera. The detector operates in pulsed mode, where it is gated to remain active only when a central slice of the Newton sphere arrives. This time-gating is achieved using a Behlke switch, capable of delivering a high-voltage pulse with a duration of 10 ns. The recorded images are analyzed offline to extract the kinetic energy (KE) and angular distribution of the ions.

## 3. Results

### 3.1. DD dynamics in $D_2$

Under the Born-Oppenheimer approximation, the molecular potential energy curves of $D_2$ are similar to those of $H_2$. Therefore, similar excitation dynamics are expected. In $D_2$, DD results in the formation of $D^-$ and $D^+$ ions upon excitation following an electron collision.

The ion yield curve of $D^-$ ions from $D_2$ shows a steady increase from 17.3 eV[13]. Upon electron impact, $D_2$ dynamics mirror those of $H_2$, with two sets of states[12] contributing to the $D^-$ signal. The first set includes Rydberg states converging to the ground state of $D_2^+$, which must survive auto-ionization and produce $D^-$ with low KE. The second set comprises neutral excited states (Q1 series) converging to the first excited state of $D_2^+$. Only ${}^1\Sigma_g^+$ and ${}^1\Sigma_u^+$ states dissociating to the $D\,(1s) + D\,(n = 2, 3, \text{and } 4)$ contribute to DD, yielding $D^-$ with significant KE above 25 eV electron energy[24] due to the extent of the Franck-Condon overlap. Reports on the absolute cross-section[12] for $D^-$ production from $D_2$ show a significant isotope effect in the DD region. This reduction is attributed to autoionization losses, as the increased reduced mass in $D_2$ results in a longer dissociation time for the molecule, leading to greater depletion through autoionization.

To obtain the effect of the isotope in quantum coherence that was observed in DD dynamics of $H_2$, VSI of $D^-$ from $D_2$ is obtained. The corresponding momentum image at 50 eV is shown in Figure 1(a). The central blob in the image corresponds to the ion-pair signal that begins to appear from the threshold, while the outer ring represents the higher KE channel of ion-pair formation. Figure 1(b) displays the corresponding KE-integrated (3 eV to 9 eV) angular distribution of the higher energy ring. This angular distribution is measured relative to the direction of the incoming electron beam. We identify this ring as the second channel contributing to ion-pair formation. The angular distribution for this channel peaks in the forward and backward directions, with a non-zero signal observed at all angles. Due to the high KE of $D^-$ and the strong Lorentz force present on the extracted ions due to the perpendicular magnetic field, there is a distortion on the left side of the VSI image. However, given the cylindrical symmetry around the electron beam and the lack of preferred orientation in the target beam molecules, we have utilized only the right half of the image for further analysis.

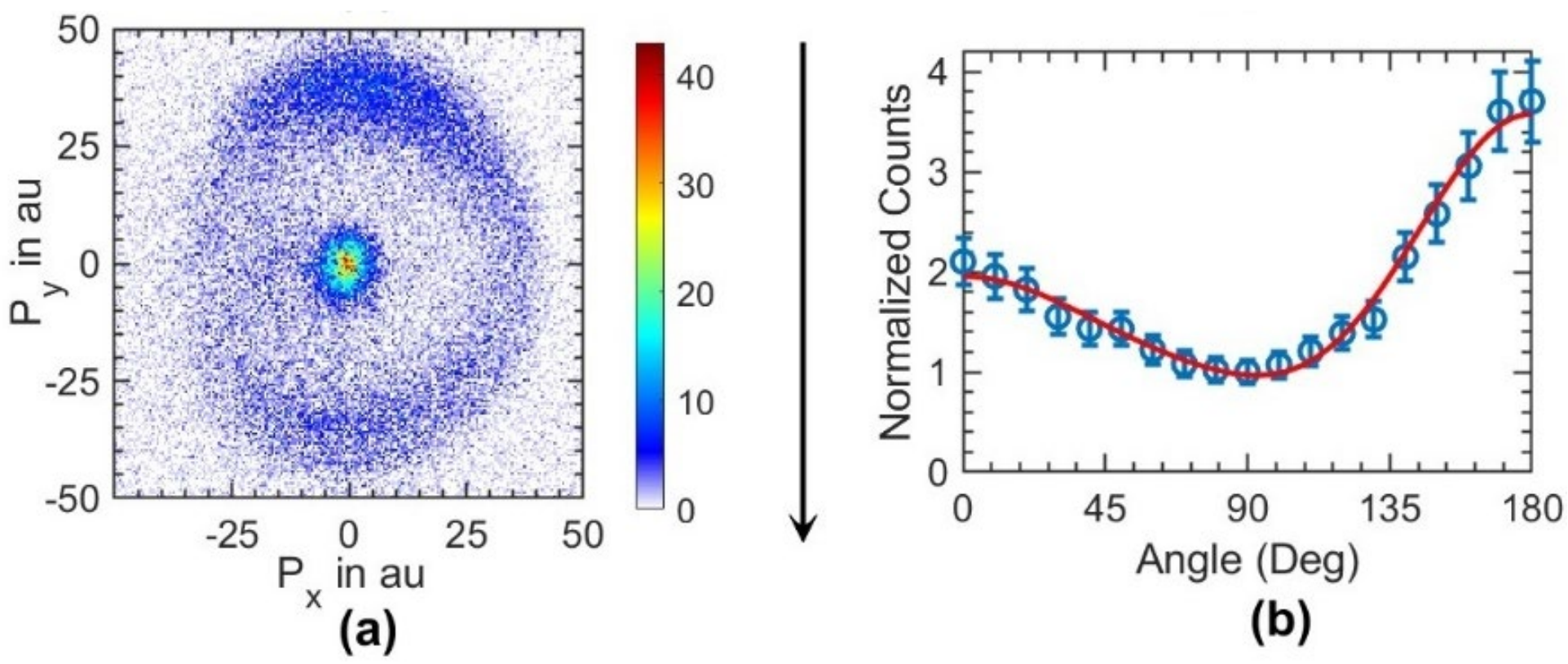

**Figure 1.** (a) Momentum image of $D^-$ from $D_2$ formed by the DD at the electron energy 50eV and the corresponding normalized angular distributions (b) obtained for the KE range 3 to 9 eV, the arrow indicates the direction of the electron beam. The corresponding fit of the angular distribution is shown by the red line obtained using Equation 2.

Similar to the $H_2$ case, the angular distribution shows unexpected forward-backward asymmetry with more intensity in the backward direction. Due to similar potential energy curves for $D_2$ as those of $H_2$, this behaviour is expected. We have also obtained the VSI of $D^-$ from $D_2$ formed by the DD due to electron scattering at different electron KE. The corresponding images also exhibit asymmetry. The observed forward-backward asymmetry in the angular distribution cannot be explained using just one of the $^1\Sigma_g^+$ or $^1\Sigma_u^+$ states. It will need a coherent superposition of the two states of opposite parities. After considering the $s$ and $d$-wave contribution in the $^1\Sigma_g^+ \rightarrow {}^1\Sigma_g^+$ transition and $p$-wave contribution to the $^1\Sigma_g^+ \rightarrow {}^1\Sigma_u^+$ transition, we obtain the fitting function for the observed angular distribution[12,27] as

$$I(\theta) = \int_{K_0-K_f}^{K_0+K_f} K^{-6} |a_g (j_0(Kr)Y_{0,0}(\theta,\varphi) + j_2(Kr)\sqrt{5}Y_{2,0}(\theta,\varphi)e^{-i\phi_1}) + a_u\sqrt{3}j_1(Kr)Y_{1,0}(\theta,\varphi)e^{-i\phi_2}|^2 KdK \quad (1)$$

where $K_0$ is the wavenumber of the incoming electron, and $K_f$ is the final wavenumber of the scattered electron. $K$ is the magnitude of momentum transfer from electron to molecule, $j_l(Kr)$ is the spherical Bessel function, and $Y_{l,\mu}(\theta,\phi)$ are the spherical harmonics. $a_g$ and $a_u$ are the effective amplitudes in each channel after the loss due to autoionization and appropriate transition to the corresponding ion-pair states. We take $\phi_1$ as the initial phase difference between the $s$ and $d$-waves, which is $\pi$. $\phi_2$ is the sum of the relative phase gained during the dissociation along the two paths and the initial relative phase between the $s$ and $p$-waves, which is $\pi/2$. Here, $r$ is related to the typical interaction length.

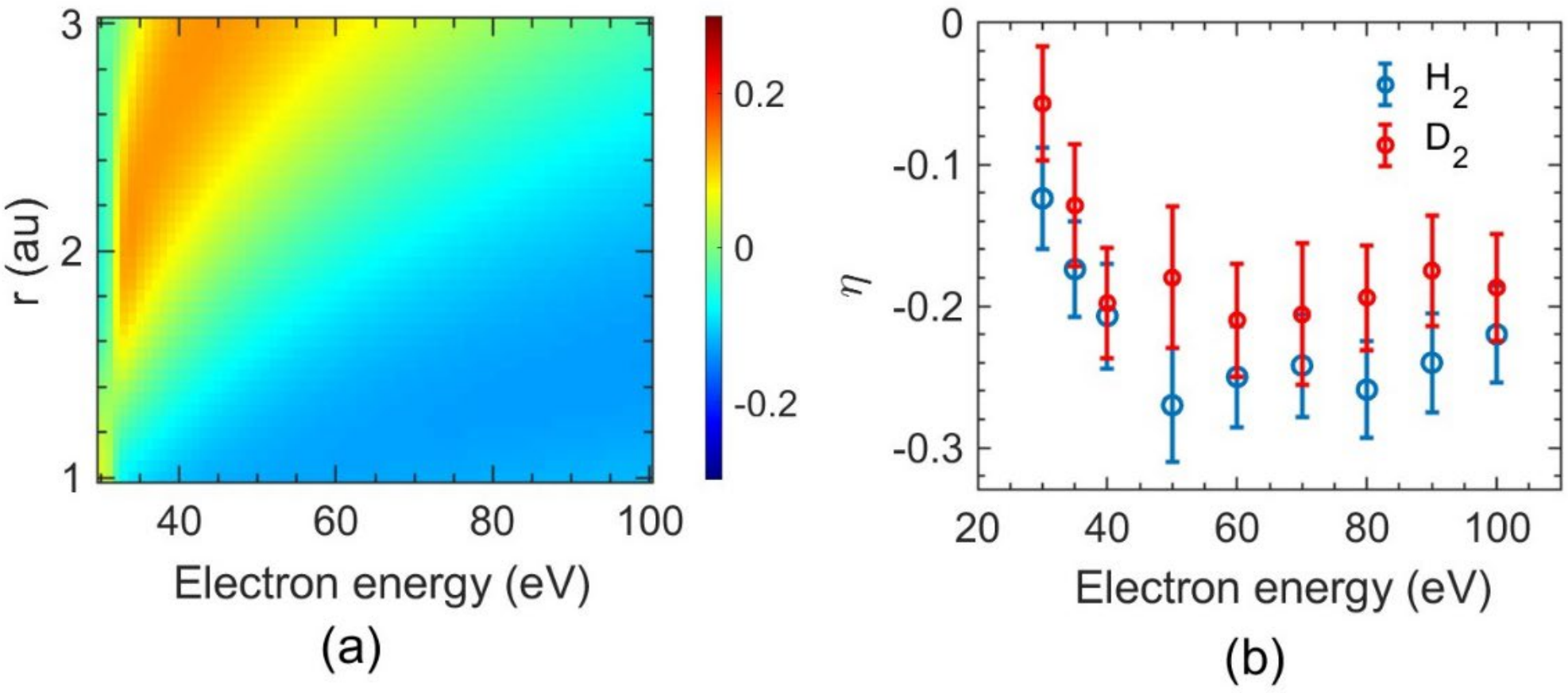

**Figure 2.** (a) The energy-integrated asymmetry parameter obtained from the model for $D_2$ as a function of incoming electron energy for various impact parameters. (b) Experimentally obtained energy-integrated asymmetry parameters for $D_2$. The corresponding values of asymmetry parameters are compared with $H_2$ (shown in blue), from ref. 12.

Using equation (2), the expected asymmetry can be calculated. We have measured the forward-backward asymmetry in terms of the asymmetry parameter $\eta$, defined as

$$\eta = \frac{I_F - I_B}{I_F + I_B} \quad (2)$$

where $I_F$ and $I_B$ are the signal intensities in the forward and backward direction with respect to the electron beam integrated over the KE range of 3 to 9 eV. The $I_F$ and $I_B$ are obtained by integrating $I(\theta)$ from equation (7) over the range 0 to $\pi/2$ and $\pi/2$ to $\pi$, respectively, and summing over the values of $K_f$. Here, parameter $r$, equivalent to the impact parameter, does not have a well-defined value. We estimate the KE-integrated asymmetry parameter over a range of values for $r$. The corresponding results are shown in Figure 2(a). The calculated results show an overall decrease in asymmetry due to the isotopic effect compared with the asymmetry estimated for $H_2$[12].

The contrast of quantum interference depends on the leftover amplitudes $a_g$ and $a_u$ of the two channels after autoionization loss and the relative phase $\phi_2$ between the two dissociating paths. Both surviving amplitudes and relative phase depend on the dissociation timescale. The surviving amplitudes vary exponentially with the dissociation time, whereas the relative phase varies linearly with it. Hence, we expect different asymmetry behavior for $D_2$. The experimentally measured values of the asymmetry parameter for $D_2$ are shown in Figure 2(b). The corresponding values of asymmetry parameters are compared with those of $H_2$ from ref. 12, shown in blue in Figure 2(b). The trend in the observed asymmetry parameters is readily apparent in simulated results, especially for impact parameters typical of the molecule's equilibrium bond length. The heavier isotope would affect the process in terms of the reduced amplitudes of the interfering wavepackets and the relative phase between them. The asymmetry parameter trend closely resembles that of $H_2$, with the main distinction being a slightly lower value observed for the asymmetry parameter compared to $H_2$.

### 3.2. DD dynamics in HD

Since, for HD, no inversion symmetry is expected, free electrons can transfer both even and odd partial waves during excitation to a given state. As we have found earlier in the case of homonuclear diatom[13] ($H_2$ and $D_2$) in the DD region, excitation to ${}^1\Sigma_g^+$ involves the transfer of $s$ and $d$ partial waves, while excitation to ${}^1\Sigma_u^+$ involves $p$ partial wave transfer. The corresponding states that are involved in excitation for HD are of ${}^1\Sigma$ symmetry. The resultant angular distribution of fragments from the ${}^1\Sigma$ state will involve $s$, $p$, and $d$ partial wave transfer of free electrons and can be expressed as

$$I(\theta) = \int_{K_0-K_f}^{K_0+K_f} K^{-6} \left| a_s j_0(Kr) Y_{0,0}(\theta,\varphi) + a_p j_1(Kr)\sqrt{3} Y_{1,0}(\theta,\varphi) e^{-i\phi_p} + a_d \sqrt{5} j_2(Kr) Y_{2,0}(\theta,\varphi) e^{-i\phi_d} \right|^2 K dK \quad (3)$$

where $K_0$ is the wavenumber of the incoming electron, and $K_f$ is the final wavenumber of the scattered electron. $K$ is the magnitude of momentum transfer from electron to molecule, $j_l(Kr)$ is the spherical Bessel function, and $Y_{l,\mu}(\theta,\phi)$ are the spherical harmonics. $a_s, a_p$, and $a_d$ are the amplitudes corresponding to $s$, $p$, and $d$ wave transfer, respectively. $\phi_p$ is the relative phase difference between the $s$ and $p$ waves. $\phi_d$ is the relative phase between the $s$ and $d$-waves. Here, $r$ is related to the typical interaction length.

Due to the difference in the electron affinities and ionization potentials of H and D atoms, there will be two dipolar dissociation limits. The two dissociation limits cannot originate from a single state; thus, there must be two contributing states. In electronic excitation, as seen in the corresponding homonuclear counterparts, two states of opposite parity ($^1\Sigma_g^+$ and $^1\Sigma_u^+$) are involved. The symmetry of the electronic states leading to the two dissociation limits in HD can be constructed using perturbation theory. For HD, a term[19] in the Hamiltonian breaks the inversion symmetry, causing a mixing of the two states of opposite parity. These resulting mixed states contribute to the dissociation limits. These linear combinations would look like

$$\psi_1 = {}^1\Sigma_g^+ + {}^1\Sigma_u^+ \quad (4a)$$

$$\psi_2 = {}^1\Sigma_g^+ - {}^1\Sigma_u^+ \quad (4b)$$

Hence, the two states, $\psi_1$ and $\psi_2$, will have an opposite phase relation between $^1\Sigma_g^+$ and $^1\Sigma_u^+$. The corresponding angular distribution from these two states would be given by

$$I_1(\theta) = \int_{K_0-K_f}^{K_0+K_f} K^{-6} |a_s j_0(Kr) Y_{0,0}(\theta,\varphi) + a_p j_1(Kr)\sqrt{3} Y_{1,0}(\theta,\varphi) e^{-i\phi_p} + a_d \sqrt{5} j_2(Kr) Y_{2,0}(\theta,\varphi) e^{-i\phi_d}|^2 K dK \quad (5a)$$

$$I_2(\theta) = \int_{K_0-K_f}^{K_0+K_f} K^{-6} |a_s j_0(Kr) Y_{0,0}(\theta,\varphi) - a_p j_1(Kr)\sqrt{3} Y_{1,0}(\theta,\varphi) e^{-i\phi_p} + a_d \sqrt{5} j_2(Kr) Y_{2,0}(\theta,\varphi) e^{-i\phi_d}|^2 K dK \quad (5b)$$

Since both states are energetically close[26] in the Franck-Condon region, $\phi p$ and $\phi d$ will have nearly the same values for both states. As a result, the cross-term in the angular distribution expression from these states will have opposite signs, leading to opposite asymmetries in the angular distribution of $H^-$ and $D^-$ if such asymmetry exists. Therefore, if HD behaves like a heteronuclear molecule, $H^-$ and $D^-$ should have complementary angular distributions.

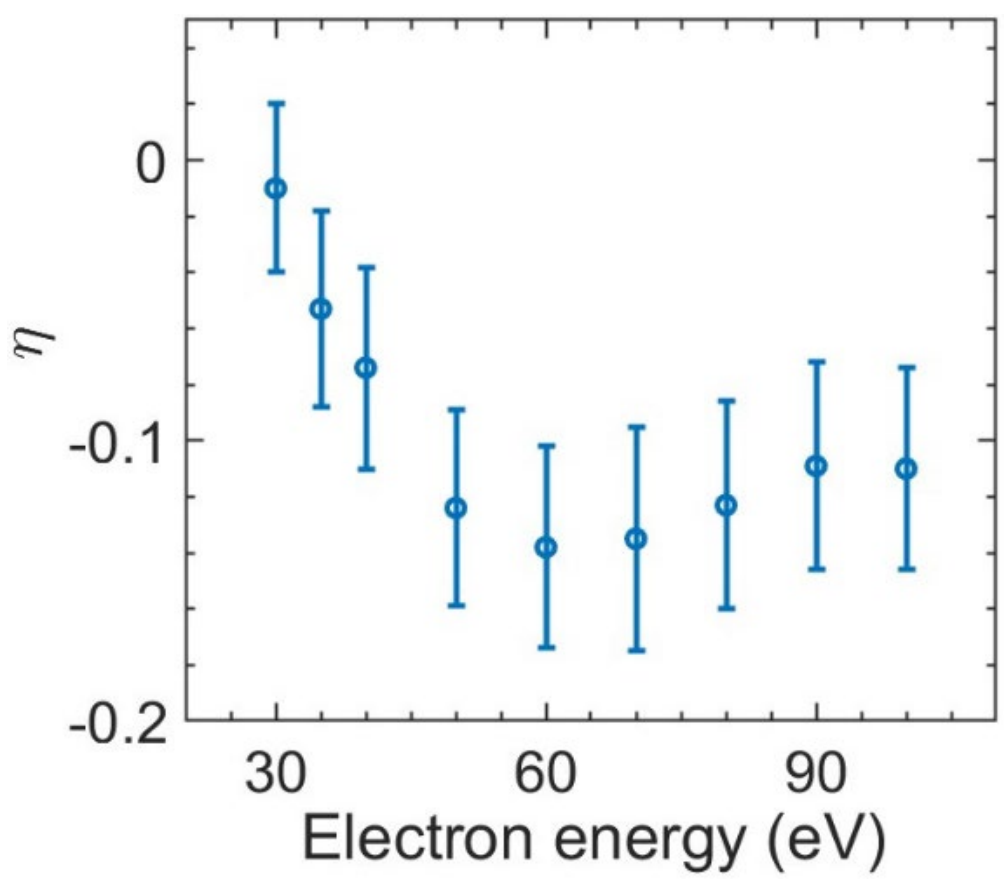

**Figure 3.** Experimentally obtained energy-integrated asymmetry parameters for $D^-$ from HD as a function of electron energy.

To examine the DD dynamics, we obtained the VSIs of $D^-$ from HD and $H^-$ from HD. Due to the asymmetric mass of HD, after dissociation, $H^-$ will have 2/3 of the KER, while $D^-$ will have 1/3 based on the momentum conservation. As a result, $H^-$ from HD has a KE of 5-12 eV. In our current experimental setup, $H^-$ ions with KE $>$ 10 eV are lost during extraction. We have obtained the VSIs of $D^-$ from HD at different incident electron energies. The data reveal a forward-backward asymmetry in the angular distribution relative to the incident electron beam. The corresponding values of the asymmetry parameter are plotted in Figure 3. As observed, the asymmetry is minimal up to 30 eV incident electron energy and begins to appear from 35 eV onwards.

If HD behaves like a heteronuclear molecule during excitation, $H^-$ should exhibit the opposite asymmetry as that of $D^-$. Despite some loss of $H^-$ signal for KE > 10 eV, the asymmetry parameter for both $H^-$ and $D^-$ can be determined within the KER range of up to 15 eV. This range corresponds to 10 eV KE for $H^-$ ions and 5 eV KE for $D^-$ ions. Since 35 eV is close to the threshold, the effect of recoil on the KE distribution will be minimal, allowing for a one-to-one correspondence between the incident electron energy and the KE of the ions, enabling us to compare asymmetry with the KE of the ions.

**Table 1.** Energy-integrated asymmetry parameters within the KER range of 15 eV at 35 eV incident electron energy for $D^-$ and $H^-$ from HD.

| Channel | $\eta$ |
|---|---|
| $D^-$from HD | -0.05 ± 0.02 |
| $H^-$ from HD | -0.06 ± 0.03 |

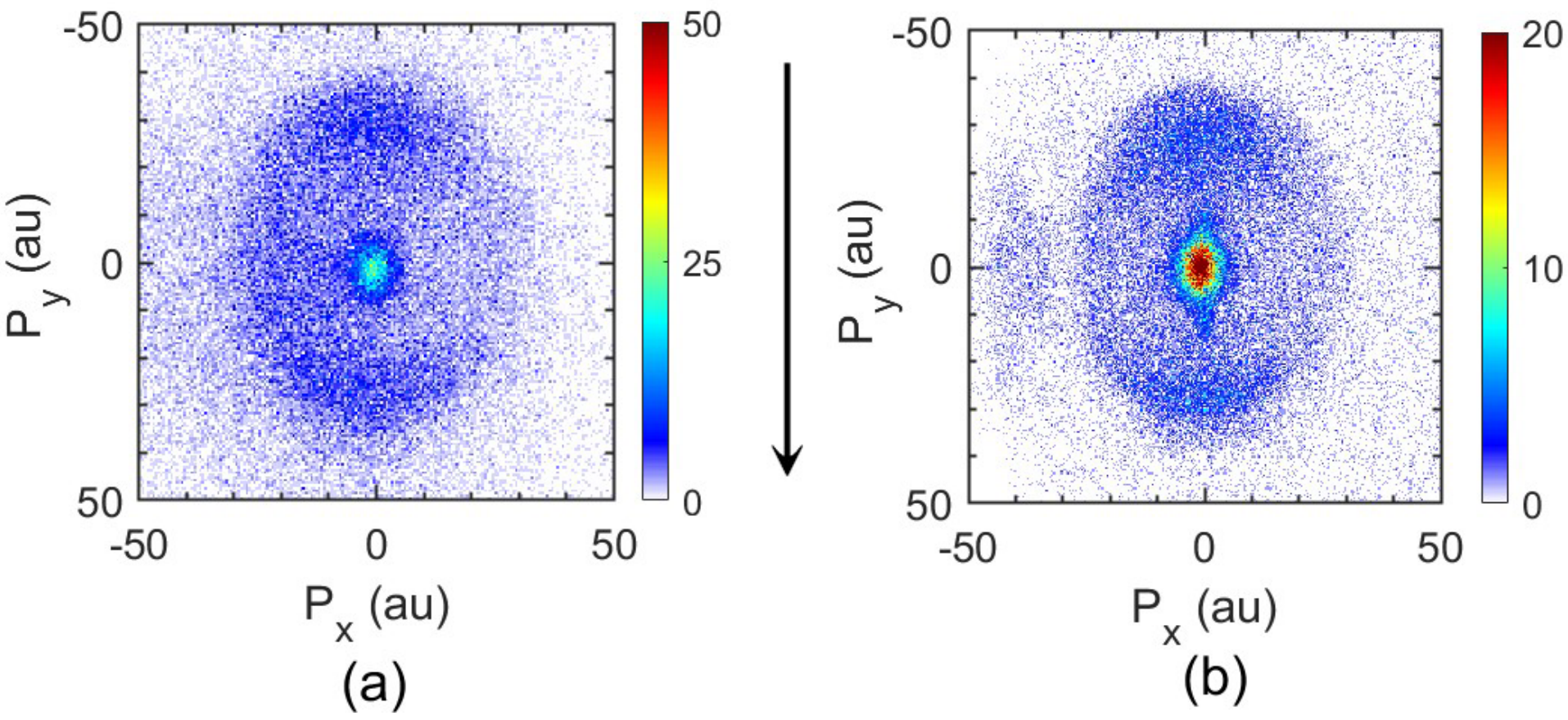

**Figure 4.** Momentum images of (a) $D^-$ and (b) $H^-$ from HD obtained at 35eV electron energy. The arrow indicates the direction of the incident electron beam.

Figure 4 compares the angular distribution of $D^-$ and $H^-$ from HD at 35 eV incident electron energy. As seen, the angular distributions of both $H^-$ and $D^-$ from HD are similar rather than complementary. The asymmetry parameters calculated for $H^-$ and $D^-$ within the KER range up to 15 eV, as shown in Table 1, were found to be similar. This finding contradicts our initial expectations, given the heteronuclear nature of HD in the Franck-Condon region. These observations are similar to those of Swain *et al.*[22], who reported identical forward-backward asymmetries for $H^-$ and $D^-$ and explained these findings in terms of the interaction of multiple resonances at the larger inter-nuclear separations. Similarly, the observed asymmetry in our study can be explained by the homonuclear-like behaviour of HD in the Franck-Condon region. The generated superposition of states interacts with other states of $^1\Sigma_g^+$ and $^1\Sigma_u^+$ symmetry at the larger inter-nuclear separations. De Lange *et al.*[19] have reported the behaviour of the $\mathrm{n} = 2$ dissociation limit for the case of HD. They noted that the potential energy curves for isotopologues of $H_2$ coincide at small internuclear separations. However, at larger internuclear distances, the potential energy curves for HD diverge from the adiabatic curve and converge to one of the dissociation limits, without crossing each other. In $H_2$ and $D_{2,}$ due to inversion symmetry, there is two-fold degeneracy. However, due to asymmetric mass, this degeneracy is lifted. Consequently, for HD, there are two distinct groups of dissociation limits for $\mathrm{n} = 2$: $\mathrm{H(2l)} + \mathrm{D(1s)}$ and $\mathrm{H(1s)} + \mathrm{D(2l)}$. Each of these limits has a near two-fold degeneracy, corresponding to $l = 0$ or $l = 1$.

If the excited atom is in the $2s$ state, the possible molecular states would be $^1\Sigma_g^+$ and $^1\Sigma_u^+$, and if the excited atom is in the $2p$ state, the possible molecular states will be $^1\Sigma_g^+$, $^1\Sigma_u^+$, $^1\Pi_g$, and $^1\Pi_u$. As $2s$ and $2p$ are near degenerate, all of these states would converge for the homonuclear case. Since the ion pair curve is of $^1\Sigma$ symmetry, only $^1\Sigma_g^+$ and $^1\Sigma_u^+$ can have avoided crossing with the ion pair curve. Out of these four states, only one pair belongs to the Q1 series in the Franck-Condon region and would contribute to the DD region corresponding to high KE ions. For the HD case, there are two distinct dissociation limits. The molecular states of heteronuclear isotopologues will be linear combinations of corresponding homonuclear states. As a result, there will be a linear combination of four $^1\Sigma$ states, two coming from $l = 2s$ and the other from $l = 2p$. These linear combinations would look like:

$$1^1\Sigma_g^+ + 1^1\Sigma_u^+ + 2^1\Sigma_g^+ + 2^1\Sigma_u^+ \quad (6a)$$

$$1^1\Sigma_g^+ + 1^1\Sigma_u^+ - 2^1\Sigma_g^+ - 2^1\Sigma_u^+ \quad (6b)$$

$$1^1\Sigma_g^+ - 1^1\Sigma_u^+ + 2^1\Sigma_g^+ - 2^1\Sigma_u^+ \quad (6c)$$

$$1^1\Sigma_g^+ - 1^1\Sigma_u^+ - 2^1\Sigma_g^+ + 2^1\Sigma_u^+ \quad (6d)$$

The prefix "1" refers to states that belong to the Q1 series in the Franck-Condon region, while the prefix "2" denotes states that are not excited in the Franck-Condon region and, therefore, not responsible for DD in the high KE region. Out of these four states, two would converge to the $H^+ + D^-$ dissociation limit, and two states would converge to the $D^+ + H^-$ limit. Based on our observation that $H^-$ and $D^-$ exhibit similar asymmetry, we propose that the combinations converge in pairs either as equations (6a) & (6c) and (6b) & (6d) or as equations (6a) & (6d) and (6b) & (6c). Among these, the states contributing to the DD signal necessarily have the same relative sign between the $1^1\Sigma_g^+$ and $1^1\Sigma_u^+$ states (i.e., 6a and 6b or 6c and 6d). This interpretation clarifies the behaviour of forward-backward asymmetry across all isotopologues. These observations also indicate that the permanent dipole moment of HD does not influence electron collisions, as observed earlier[21,22] in DEA experiments.

Interestingly, a comparison between the asymmetry parameters for $D^-$ from HD and that for $H^-$from $H_2$, and $D^-$from $D_2$ shows that the asymmetry observed in HD is lower than the other two isotopologues at any given electron energy. One possible explanation for this can be based on the varying phases and amplitudes between the two paths due to different timescales of dissociation. However, as the overall asymmetry parameter does not change drastically over the observed electron energy range, this may not provide the correct trend. For example, the dissociation time for HD must be between $H_2$ and $D_2$, the loss due to autoionization should also show a similar trend, making it maximum for $D_2$ and minimum for $H_2$. In the electron capture experiment on $HD^+$, it has been shown that the resulting DD shows quantum interference arising from the dissociation through multiple potential energy curves crossing each other[28]. Such curve crossings will be more in HD than in homonuclear counterparts due to the loss of inversion symmetry as the internuclear separation increases. This can result in lower contrast in HD compared to the homonuclear counterparts.

Here, we have considered only the n = 2 state and formed a linear combination to describe the observed dynamics. However, states with n = 3 and n = 4 also contribute to the DD dynamics. In reality, many $^1\Sigma$ states are involved in the DD dynamics. Moreover, the DD process can, in principle, result from multiple curve crossings during dissociation among the neutral excited states[28]. Hence, there is a critical necessity for precise calculations using modern theoretical tools to accurately describe the dynamics of these simple diatomic molecules in terms of coherent superpositions of states.

## 4. Conclusion

To summarize, our findings highlight the isotope effect in quantum coherence triggered by non-resonant inelastic electron scattering in molecular dynamics. For $D_2$, we have obtained the VSIs of $D^-$ at different electron energies. From these VSIs, we have identified an asymmetry in the angular distribution. The asymmetry parameter for $D^-$ from $D_2$ exhibits a similar pattern to that of $H^-$ from $H_2$, with the only distinction being a slightly lower value for $D_2$. We attribute this to the loss of contrast in the quantum interference due to slower dissociation.

In the case of HD, we have observed similar angular distributions for $H^-$ and $D^-$. This finding contradicts the expected behaviour if HD were acting as a heteronuclear molecule in the Franck-Condon region. We have explained the identical asymmetry of $H^-$ and $D^-$ in terms of homonuclear-like behaviour in the Franck-Condon region. Excitation in the Franck-Condon region prepares a coherent superposition of states, similar to what occurs in homonuclear counterparts. At large internuclear separations, this superposition interacts with other states, resulting in identical forward-backward asymmetry for both $H^-$ and $D^-$. These findings indicate that the permanent dipole moment of HD does not play a significant role in electron-impact excitation, as observed in electron-capture processes.

**Acknowledgment**

The authors acknowledge the financial support from the Department of Atomic Energy, India, under Project Identification No. RTI4012.

## References

[1] Christophorou L G 1983 *Electron-Molecule Interactions and Their Application. Vol. 1* (Academic Press, London) (DOI: 10.1016/B978-0-12-174401-4.X5001-9)

[2] Ingólfsson O 2019 *Low-Energy Electrons Fundamentals and Applications* (Jenny Stanford Publishing, New York) (DOI: 10.1201/9780429058820)

[3] van Dorp W F 2016 *Theory: Electron-Induced Chemistry*. *Frontiers of Nanoscience* **11** 115 (Elsevier) (DOI: 10.1016/B978-0-08-100354-1.00003-X)

[4] van Dorp W F 2012 The role of electron scattering in electron-induced surface chemistry *Phys. Chem. Chem. Phys.* **14** 16753 (DOI: 10.1039/C2CP42275A)

[5] Zhong Q, Ihle A, Ahles S, Wegner H A, Schirmeisen A, and Ebeling D 2021 Constructing covalent organic nanoarchitectures molecule by molecule via scanning probe manipulation *Nat. Chem.* **13** 1133 (DOI: 10.1038/s41557-021-00773-4)

[6] Nandi D, Prabhudesai V S, Krishnakumar E, and Chatterjee A 2005 Velocity slice imaging for dissociative electron attachment *Rev. Sci. Instrum.* **76** 053107 (DOI: 10.1063/1.1899404)

[7] Adaniya H, Rudek B, Osipov T, Haxton D J, Weber T, Rescigno T N, McCurdy C W, and Belkacem A 2009 Imaging the Molecular Dynamics of Dissociative Electron Attachment to Water *Phys. Rev. Lett.* **103**, 233201 (DOI: 10.1103/PhysRevLett.103.233201)

[8] Slaughter D S, Adaniya H, Rescigno T N, Haxton D J, Orel A E, McCurdy C W, and Belkacem A 2011 Dissociative electron attachment to carbon dioxide via the 8.2 eV Feshbach resonance *J. Phys. B: At. Mol. Opt. Phys.* **44** 205203 (DOI: 10.1088/0953-4075/44/20/205203)

[9] Bingham P S and Gorfinkiel J D 2024 Resonances in electron scattering from $H_2$ around the H(2l) + $H^-(1s^2)$ dissociation limit *J. Phys. B: At. Mol. Opt. Phys.* **57** 105202 (DOI: 10.1088/1361-6455/ad38f3)

[10] Ambalampitiya H B, and Fabrikant I I 2026 Dissociative electron attachment to $H_2$: two resonances of different symmetries *Phys. Re. A* **113** 012811 (DOI: 10.1103/f33t-9lk9)

[11] Krishnakumar E, Prabhudesai V S, and Mason N J 2018 Symmetry breaking by quantum coherence in single electron attachment *Nat. Phys.* **14** 149 (DOI: 10.1038/nphys4289)

[12] Kumar A, Swain S, and Prabhudesai V S 2023 Inelastic electron scattering induced quantum coherence in molecular dynamics *Nat. Commun.* **14** 2769 (DOI: 10.1038/s41467-023-38440-6)

[13] Krishnakumar E, Denifl S, Cadez I, Markelj S, and Mason N J 2011 Dissociative Electron Attachment Cross Sections for $H_2$ and $D_2$ *Phys. Rev. Lett.* **106** 243201 (DOI: 10.1103/PhysRevLett.106.243201)

[14] Prabhudesai V S, Mason N J, and Krishnakumar E 2020 Electron attachment and quantum coherence in molecular hydrogen *J. Phys. Conf. Ser.* **1412** 052006 (DOI: 10.1088/1742-6596/1412/5/052006)

[15] Blinder S M 1960 Dipole moment of HD *J. Chem. Phys.* **32** 105 (DOI: 10.1063/1.1700880)

[16] Herzberg G 1950 Rotation-vibration spectrum of the HD molecule *Nature* **166** 563 (DOI: 10.1038/166563a0)

[17] Biesheuvel J, Karr J-Ph, Hilico L, Eikema K S E, Ubachs W, and Koelemeij J C J 2016 Probing QED and fundamental constants through laser spectroscopy of vibrational transitions in $HD^+$ *Nat. Commun.* **7** 10385 (DOI: 10.1038/ncomms10385)

[18] Fichet S 2018 Quantum Forces from Dark Matter and Where to Find Them *Phys. Rev. Lett.* **120** 131801 (DOI: 10.1103/PhysRevLett.120.131801)

[19] De Lange A, Reinhold E, and Ubachs W 2002 Phenomena of *g-u* symmetry-breakdown in HD *Int. Rev. Phys. Chem.* **21** 257 (DOI: 10.1080/01442350210124515)

[20] Sheehy B, Walker B, and DiMauro L F 1995 Phase Control in the Two-Color Photodissociation of $HD^+$ *Phys. Rev. Lett.* **74** 4799 (DOI: 10.1103/PhysRevLett.74.4799)

[21] Swain S, Krishnakumar E, and Prabhudesai V S 2021 Structure and dynamics of the negative-ion resonance in $H_2$, $D_2$, and HD at 10 eV *Phys. Rev. A* **103** 062804 (DOI: 10.1103/PhysRevA.103.062804)

[22] Swain S, Kumar A, Krishnakumar E, and Prabhudesai V S 2024 Quantum coherence in dissociative electron attachment: Isotope effect *Phys. Rev. A* **109** 042805 (DOI: 10.1103/PhysRevA.109.042805)

[23] https://webbook.nist.gov/chemistry/.

[24] Kumar A, Swain S, Upadhyay J, Upalekar Y, Arya R, and Prabhudesai V S 2023 Predissociation dynamics of negative-ion resonances of $H_2$ near 12 and 14.5 eV using the velocity slice imaging technique *Phys. Rev. A* **107** 062803 (DOI: 10.1103/PhysRevA.107.062803)

[25] Das S, Swain S, Gope K, Tadsare V, and Prabhudesai V S 2024 Effect of static gas background signal on momentum imaging in electron-molecule collision experiment *Int. J. Mass. Spectrom.* **498** 117215 (DOI: 10.1016/j.ijms.2024.117215)

[26] Vogel J S 2015 $H^-$ formation by neutral resonant ionization of H($n$=2) atoms *AIP Conf. Proc.* **1655** 020015 (DOI: 10.1063/1.4916424)

[27] Van Brunt R J 1974 Breakdown of the dipole-Born approximation for predicting angular distributions of dissociation fragments *J. Chem. Phys.* **60** 3064 (DOI: 10.1063/1.1681491)

[28] Zong W *et al*. 1999 Resonant ion pair formation in electron collision with ground state molecular ions *Phys. Rev. Lett.* **88** 951 (DOI: 10.1103/PhysRevLett.83.951)